# Architecture for Automated Tagging and Clustering of Song Files According to Mood

Puneet SINGH[1], Ashutosh KAPOOR[2], Vishal KAUSHIK[3] and Hima Bindu MARINGANTI [4]

[1,2,3,4] Jaypee Institute of Information Technology
NOIDA, Uttar Pradesh, India

**Abstract**

Music is one of the basic human needs for recreation and entertainment. As song files are digitalized now a days, and digital libraries are expanding continuously, which makes it difficult to recall a song. Thus need of a new classification system other than genre is very obvious and mood based classification system serves the purpose very well.
In this paper we will present a well-defined architecture to classify songs into different mood-based categories, using audio content analysis, affective value of song lyrics to map a song onto a psychological-based emotion space and information from online sources. In audio content analysis we will use music features such as intensity, timbre and rhythm including their sub-features to map music in a 2-Dimensional emotional space. In lyric based classification 1-Dimensional emotional space is used. Both the results are merged onto a 2-Dimensional emotional space, which will classify song into a particular mood category. Finally clusters of mood based song files are formed and arranged according to data acquired from various Internet sources.

## 1. Introduction

Listening music is one of the oldest and the easiest way to entertain oneself and change one's mood. But as the digital libraries are increasing with myriad songs coming each year, it is becoming difficult for listeners to keep track of each song. There are several problems faced by listeners when they have the high freedom of choice as they favor upper 20% of the items against the other 80%; this is known as 80-20 rule (Pareto Principle). Thus there is a need for a system which retrieves and recommends as well as arranges music according to a simple yet important category i.e., mood.

A song comprises of two very important components, music and lyrics. Thus the effect of both on the psychology is the basis of the work presented here. A vast research has already been done in the field of music information retrieval and relation between music and emotion; as a result there exists many different models of human emotions and their relation with music. The objective of the architecture is to use best possible psychological model of emotion and to incorporate the findings of these studies in the best possible way so that music can be queried, browsed or explored by mood, rather than by artist, album or genre.

The very basic approach of the work presented here are the features extracted from an audio file (WAV format) which are used in combination with the affective value of song lyrics and other data from Internet to map a song onto a psychological-based emotion space. First the audio content of the song is analyzed and intensity, timbre and rhythm features are extracted, then lyrics file is analyzed using language processing. Then the mathematical values of all these data are mapped onto 2-Dimensional emotional space (Thayer's Energy Stress Model), each quadrant representing a set of moods. Finally play-lists are generated for these four quadrants which are afterwards arranged according to the data extracted from Internet.

Such a system can have the following uses:
- Context-Aware Play-list Generation
- Retrieving Lost Music
- Music Classification
- Music Recommendation.

### 1.1 Mood Models

To classify a song based on mood, it is really necessary to choose a model to map mathematical values onto an emotional space so that we can separate each song from another according to mood. There are many emotional models[10][11][12][13] in existence and traditionally they all use adjective descriptor based approach such as 'gloomy', 'depressing', 'energetic' etc. But the problem with such an approach is the wide range of adjective which keeps on changing in different mood models thus there is no standard model in this type of approach. Furthermore, the more adjectives there are there the more would be the complexity and probability of error in mood detection. This ambiguity makes it difficult to detect the true mood, hence a simpler and yet a profound model was needed and work by Thayer[9][13] is best for this purpose.

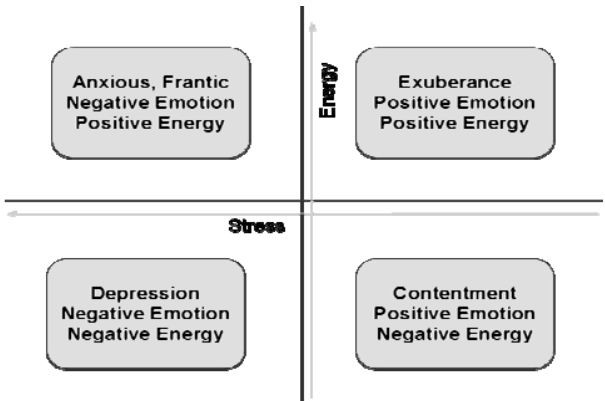

Figure. 1 Thayer's Model of mood

Unlike other adjective based models which collectively form a mood pattern, this dimensional approach adopts the theory that mood is entailed from two factors:
- Stress (happy/anxious)
- Energy (calm/ energetic)

And divides music mood into four clusters: Contentment, Depression, Exuberance and Anxious/Frantic as shown in Fig.1.

## 2. Mood detection using music

In this section we will discuss the algorithm, design and complete approach of mood detection using audio content analysis. Detection of mood from music needs a very basic step of retrieving information from music which can be done by extracting features such as intensity, timbre and rhythm which have their sub features. For extracting features, first all the music files are down-sampled to 16000Hz, 16 bits, and mono-channel. Then a set of 10 features are extracted from music for every 10 seconds of the music file. Once all the features are extracted for a song on every 10 second basis then all features are normalized using Max-Min normalization so that each song has the uniform scale.

$$v' = \frac{v - \min(v)}{\max(v) - \min(v)} \times 100 \qquad (1)$$

Thus each song will have values of every feature in between 0 to 100, for every 10 seconds of the song. After normalizing sub features, the overall parent feature value is calculated using weighted average of sub features.

At last these mathematical values are scaled on a scale of -10 to 10 making it easy to map the features on Thayer's mood model.

2.1 Feature extraction

Feature extraction is necessary for knowing the type of music, It was indicated that mode, intensity, timbre and rhythm are of great significance in arousing different music moods (Hevner, 1935; Radocy, and Boyle, 1988; Krumhansl, 2002).We have used only three features viz intensity, timbre and rhythm to map onto Thayer's mood model, where intensity is associated with energy while timbre and rhythm are in combination associated with stress.

**Intensity:** This feature has two sub-features: signal's RMS and low-energy. RMS of a signal is a more prominent feature out of the two. If intensity is higher than it can be a song with excitement or a frantic song. Low-energy of a signal is the average of the values of the peak having values lower than a threshold which is generally very low. Thus low-energy will have values opposite to RMS i.e., for a song having high RMS, value of low-energy would be low. Overall intensity is calculated using weighted average of the two sub-features:

*Intensity = (RMS) × 0.8 + (100-lowenergy) × 0.2* (2)

Low-energy is subtracted from 100 because using Eq. (1) we have scaled low-energy on a scale of 0-100.
This overall Intensity is extracted for each 10 second of the song file.

**Timbre:** Timbre is the quality of the music. Existing results show that the timbre of sound is determined primarily by the spectral information [16].
We have used four features for extracting timbre:

- Zero-cross (Z)
- Centroid (C)
- Roll off (R)
- Entropy (E)

Higher values of the feature mean lower stress on Thayer's 2-D emotional space. Thus we can say that a higher centroid is generally associated with positive emotion.
Centroid and roll off are major source of mapping emotional intensity out of these four features.
Overall timbre is calculated using weighted average of these four features:

*Timbre = Z × 0.2 + C × 0.4 + R × 0.3 + E × 0.1* (3)

**Rhythm:** It is said that three features of rhythm are more important for music strength, tempo and regularity [1].
Thus, more tempo and regularity in the song means positive emotion while low rhythmic values will mean negative emotion.
The following features are extracted for calculating rhythm:

- Average Beat-spectrum (B)
- Average Tempo (T)
- Average Regularity (Ar)

- Average Positive Autocorrelation (Pa)
- Average Negative Autocorrelation (Na)

Out of the five features most important features are tempo, beat-spectrum and regularity.

Overall rhythm is calculated using weighted average of these five features:

$Rhythm = B \times 0.25 + T \times 0.45 + Ar \times 0.2 + (Pa + Na) \times 0.05$ (4)

### 2.2 Scaling and Averaging

After calculating features for every 10 seconds of the song file we have generated arrays of these three features. Now scaling is done to map the values of the features on a scale of (-10, 10). Scaling is important for mapping the values retrieved on Thayer's mood model. Before scaling we have passed each value of feature into the following equation:

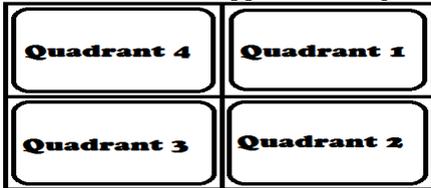 (5)

By passing through this equation we have increased each value by a factor of the maximum value making the maximum value to be 100%.

After scaling, an average is taken of all the scaled values for each feature, the result would be the final mathematical values of the all three features of the song.

### 2.3 Mood Mapping

This will be the final and important step of mood detection. As we have discussed earlier that the mathematical values are mapped onto 4 quadrants:

| Quadrant 4 | Quadrant 1 |
| Quadrant 3 | Quadrant 2 |

Figure. 2 The four quadrants of the Thayer's mood model

Table 1: Valance - Arousal relation with Thayer's Model

| Quadrant (Figure 2) | Valance (Emotion) | Arousal (Intensity) | Possible Moods |
|---|---|---|---|
| 1 | Positive | Positive | Very happy, Exciting |
| 2 | Positive | Negative | Soothing, Pleasure |
| 3 | Negative | Negative | Sad, Depressing |
| 4 | Negative | Positive | Frantic, Tense |
| 1 and 2 | Positive | zero | Serene, Happy |
| 2 and 3 | Zero | Negative | Dreamy, Sad |
| 3 and 4 | Negative | Zero | Sorrow, Disturbing |
| 1 and 4 | Zero | Positive | Exciting, Disturbing |
| 1, 2, 3 & 4 | Zero | Zero | Nothing can be said |

When the features are scaled on the scale of (-10,10) then the feature having value in between (-1,1) will be marked as 'Zero'. The feature been marked as 'Zero' can be on either side of the quadrant. For calculating Valence out of timbre and rhythm, it has been stated that Timbre features are more important than the Rhythmic features for differentiating between contentment and depression while vice-versa is true for exuberance and anxious differentiation [1]. Thus to calculate valence (emotion) of the song we have used the following equation:

$Valence [2 – 3] = Timbre \times 0.7 + Rhythm \times 0.3$ (6)

$Valence [1 – 4] = Timbre \times 0.3 + Rhythm \times 0.7$ (7)

### 2.4 Results

Using this process we have completed the first step of mood detection of song file, we have tested the algorithm on more than 70 Indian Songs and reached an accuracy of 70.7%. The confusion matrix is as follows:

Table 2: Mood Detection confusion matrix

| Quadrant | 1 | 2 | 3 | 4 |
|---|---|---|---|---|
| 1 | **75.6** | 6.3 | 0 | 18.1 |
| 2 | 6.3 | **65.7** | 24.1 | 3.9 |
| 3 | 0 | 20.6 | **70.1** | 9.3 |
| 4 | 14.2 | 0 | 8.4 | **77.4** |

As we aimed to tag a song file which not only contains music but also contains singer's voice which adds the confusion and ambiguity resulting lesser accuracy. Also the Indian music is too versatile, for example 'Sufi' songs bring contentment but are tagged as energetic song as they have high rhythm and timbre as well as intensity.

Mood of a song is also subjective to the listeners; hence the accuracy varies from listener to listener. Thus we can say that the results obtained are satisfactory.

## 3. Mood detection using Lyrics

In this section, complete algorithm and approach we have used for mood detection using lyrics has been discussed. Our aim is to refine the results of audio content analysis; we examined previous works which used the same approach [14] [15], and found that lyrics are a weak source of mood information. As the lyrics do not follow any particular grammar rule in favor of making it more rhyming, this weak source can considerably improve the result coming from music analysis. Meaning extraction from lyrics is a tedious task thus we used Bag-of-Words technique combined with Bi-gram model. Both the techniques are not based on semantic analysis of the text; these techniques are dependent on adjectives and combination of two adjectives.

Unlike the mood mapping technique used earlier [14] we have observed that lyrics have more associativity with valence than arousal. When we analyzed 'lrc' files which have timestamp of each lyric phrase sung and plotted a graph with respect to time with the arousal and valence, we got the following results:

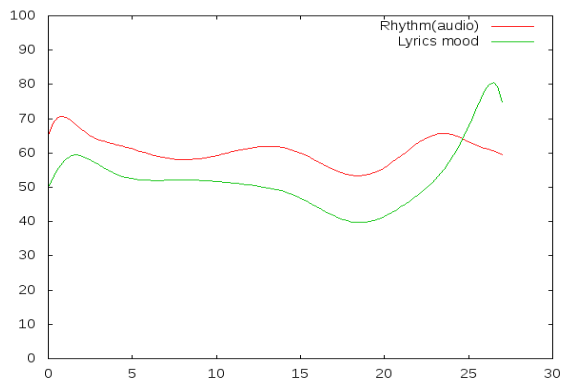

Figure. 3 Rhythm and lyrics analysis graph for the song Roobaroo( Rang de basanti) which depicts the associativity of this song's lyrics with rhythm.

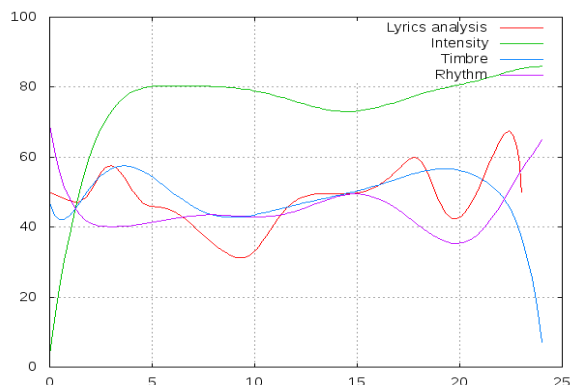

Figure. 4 This plot clearly depicts the associativity of lyrics analysis with valence. (The song is chor bazari from the movie "Love aaj kal")

Figure 4 clearly depict the disassociation of results of lyrics analysis with the intensity feature extracted. This was true for most of the songs we analyzed. This motivated us to use 1-D model for lyrics analysis rather than a 2-D model. In this 1-Dimensional model a lyrics are classified in two categories:
- Positive Emotional
- Negative Emotional

The other reason to use this approach was the fact that classification on the basis of intensity has much more accuracy than on the basis of valence. In Table 2 the confusion matrix has lower confusion in 1-2, 1-3, 2-1, 2-4, 3-1, 3-4, 4-2, 4-3. Hence we can say that if we can reduce the confusion between 1-4, 4-1, 2-3 and 3-2 it will improve the accuracy of the results. Thus we do not have much need of a two dimensional model for refining the results, and mapping lyrics emotion onto a 1-D model is less complex than onto the 2-D.

3.1 Bag-of-Words Analysis

Bag-of-Words or BOW is a collection of words where each word is assigned tags from a dictionary. A word can have different tags, each tag is predefined as positive or negative according to mood.

We have used the following tags:

Table 3: Different tags used for BOW technique

| Positive Emotion | Negative Emotion |
|---|---|
| Happy words | Sad words |
| Sexual Words | Anger Words |
| Insightful words | Death Words |
| Achievement words | |

Other tags which we have decided on the basis of previous tags were:
- Social words
- Feel
- Family
- Health
- Affect words
- Friendship
- Body
- Cause
- Inclusion
- Exclusion

We first calculated all positive and all negative words using simple word count, and then the left out tags were tagged on the basis of the values attained from the positive and negative word counts.

The left out tags are used according to the following rules:

if(total positive > 1.5 X total negative)
{
total positive = total positive + Social + Feel + Family + Friend + Affect + Cause + Inclusion + Exclusion + Body + Health;
}
else if(total negative > 1.5 X total positive)
{
total negative = total negative + Social + Feel + Family + Friend + Affect + Cause + Inclusion + Exclusion + Body + Health;
}

Thus the left out tags will be added only in the cases when positive and negative tags are 1.5 times the other. This will help in classifying the lyrics in three major categories High, Medium and Low Positive Emotion. The final emotion would be in terms of percentage of positive and negative emotions.

If positive emotion in lyrics is more, clearly it means that song contains more words tagged as positive on emotional scale. The final result will be on a scale of 0 to 100.

### 3.2 Bi-gram Analysis

Bi-gram analysis can be said to work as a pseudo-semantic analysis where no actual meaning is extracted but still a meaningful emotional value is generated. We used the same tags as of BOW but analyzed the combination of the two words.

We marked each combination with a certain value lying between the range (-10, 10) depending on the mood they will elicit when combined together. There are more than 100 combinations possible for analysis; a few amongst them are as follow:

Table 4: Bi-gram showing different emotional values

| Tag1 | Tag2 | Emotional value | Example |
|---|---|---|---|
| Happy | Family | 6 | Smiling Brothers |
| Exclusion | Family | -5 | Without Brothers |
| Body | Sexual | 4 | Hand Massaged |
| Negate | Happy | -6 | Not Glad |
| Cause | Happy | 7 | Made Glad |

After marking each bi-gram with a mathematical value we again calculated positive and negative percentages just as we did in BOW technique. We could have used trigram or n-gram for better results, but as the number of combinations increase the complexity increases. Bi-gram itself had more than 100 combinations, thus mood detection on the basis of the above mentioned two factors is optimal in terms of complexity.

### 3.3 Weighted average and mood mapping

A weighted average is taken of the values attained from both techniques.
As it is said that BOW is an equally important technique, so the weighted average is taken as:

$Emotion = BOW \times 0.4 + Bigram \times 0.6$ (8)

If finally the value of positive emotion is above 60 % then we will say the song has positive valence, and if the final emotional value is below 40 % we will say the song havs negative valence.

We have not used the extent of positivity or negativity to refine the result; if the song was lyrically positive we added 1 to the valence and subtracted 1 if the song was negative. Thus on a 1-D emotional model we mapped any song file as positive emotional, negative emotional and if the value lies between 40-60 % then results could not be refined on the basis of lyrics.

### 3.4 Results

Using this process we have refined the results significantly. The confusion matrix of the result from lyrics analysis is as follows:

Table 5: Lyrics based mood detection confusion matrix

| Emotion | Positive | Negative | NULL |
|---|---|---|---|
| Positive | **62.3** | 7.3 | 30.4 |
| Negative | 9 | **64.4** | 26.6 |
| NULL | 12.2 | 8.5 | **79.3** |

As the NULL result will not affect the results of the audio content analysis, and the confusion in positive and negative is lesser so we can assume that Lyrics analysis method will refine the result without harming it much. We will discuss the final confusion matrix and accuracies in Results (5) section.

## 4. Implementation and Architecture

Implementation is one of the important tasks to support research; we accomplished this task by using best tools available for each problem. We used Matlab for extracting music features using MIRToolBox, Perl for language processing of lyrics, and finally used the both results, generated clusters (play list of a particular mood) and sorted them according data acquired from Internet in C#.

We have used XML file to store the data acquired from audio and lyrics processing and used them from C#. The basic idea behind the use of XML file is that it can store live data; using XML we can create a system where listeners can directly download these preprocessed files from Internet and use a music player to sort their music according to mood with a mood equalizer. The XML file will have all the data needed to support the mood detection with some meta-data and manually tagged genre of song file.

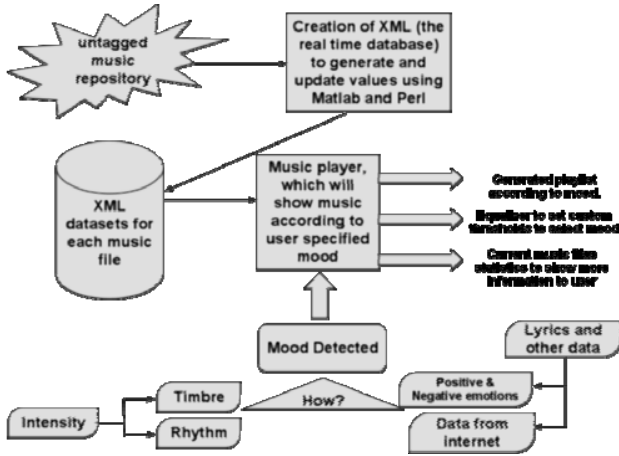

Figure. 5 Complete architecture of the implementation

After mood detection is complete we have used You-tube's like dislike factor and IMDB rating of the album for sorting the clusters.

## 5. Results and Discussion

After the whole processing we will have 4 clusters (play-lists) as a result arranged in order of probability of likeness.
The final confusion matrix of the final result is as follows:

Table 6: Final Mood Detection confusion matrix

| Quadrant | 1 | 2 | 3 | 4 |
|---|---|---|---|---|
| 1 | **83.2** | 6.3 | 0 | 10.5 |
| 2 | 6.3 | **79.4** | 10.4 | 3.9 |
| 3 | 0 | 13.4 | **77.4** | 9.2 |
| 4 | 10.5 | 0 | 8.4 | **81.1** |

This clearly indicates the reduction in confusion and ambiguity. The overall accuracy after lyrics analysis refining, is increased approximately to 79.2 %. It must be noted that for attaining more efficiency we must have songs of more than 1 minute duration which will give it at least 6 different feature sets for better normalization. Corrupted of pirated music files having high noise ratio will also face the problem as noise will add up to the RMS values and disturbing timbre and rhythmic values.
The data set we have used for the experiment was Indian music which was the basis of the whole research. Indian music has lyrics in 'Hindi' language so we converted each 'lrc' file into English, because English corpora are easily available. But the results will not be degraded using this process because our technique was not based on semantic analysis but the bag of words and word combinations, and adjectives and tags generally remains the same in Hindi as they were in English.
After mood detection we arranged the clusters using the data we acquired from You-tube and IMDB. Suppose a song has 98 likes and 2 dislikes and has IMDB rating of the album 7.9 out of 10 while another in the same cluster has You-tube's 93 likes and 7 dislikes and IMDB rating for the album to be 8.5 out of 10. So we simply multiplied "likeness factor's percentage" with IMDB rating.

Table 7: Calculating likeness factor of two songs

| Song | Total rating |
|---|---|
| Song 1 | 98% X 7.9 = 7.742 |
| Song 2 | 93 % X 8.5 = 7.905 |

Thus Song 2 will have more likeness factor hence is arranged above Song 1 while play list generation.

## 6. Conclusion and future work

In this article we discussed simpler but detailed algorithms to find mood from a song file. The architecture is not completely new but the way of implementation is novel. The results we obtained from experiment and implementation were more than satisfactory. The basic motive behind the experiment was to create a system where music listeners can browse music according to mood of the music. In future one can try to improve the results using better music feature extraction and incorporating better lyric analysis methods, music features are really important for the type of music. Thus research can be done to find the best features for a particular type of music. Internet digital libraries are continuously tagged by millions of users for example www.last.fm can be used to find genre and finding features according to the genre. This could be a hierarchical model where at each stage results are generated at each node and classified.

### Acknowledgments

We would like to acknowledge all the people who participated in the survey to tag the music in a particular quadrant so that we can validate our result. We would specially like to thank Prabodh Prakash and Tanvi Shukla for continuously motivating and helping to make the work possible. Finally we would like to thank Department of Computer science of Jaypee Institute of Information Technology for providing us the platform and opportunity to work in this area.